\documentclass[prd,aps,showpacs,preprintnumbers,amssymb]{revtex4}

\usepackage{epsf}

\begin{document}

\title{%
\hfill{\normalsize\vbox{%
\hbox{\rm SU-4252-879}
}}\\
{
Exploration of a physical picture for the QCD scalar channel
}}

\author{Amir H. Fariborz $^{\it \bf a}$~\footnote[3]{Email:
fariboa@sunyit.edu}}

\author{Renata Jora $^{\it \bf b}$~\footnote[2]{Email:
cjora@physics.syr.edu}}

\author{Joseph Schechter $^{\it \bf 
c}$~\footnote[4]{Speaker;Email:
schechte@physics.syr.edu}}

\affiliation{$^ {\bf \it a}$ Department of Mathematics/Science,
State University of New York Institute of Technology, Utica,
NY 13504-3050, USA.}

\affiliation{$^ {\bf \it b,c}$ Department of Physics,
Syracuse University, Syracuse, NY 13244-1130, USA,}

\date{\today}

\begin{abstract}       
   A generalized linear sigma model 
is employed to study the quark structure of
 low lying scalar as well as pseudoscalar states.
The model allows the possible mixing of
 quark anti-quark states
with others made of two quarks and two antiquarks but
 no {\it a priori} assumption is made about the
 quark contents of the predicted physical states.
Effects of SU(3) symmetry breaking are included. 
The lighter conventional pseudoscalars turn out to be
 primarily of
two quark type whereas the lighter scalars
turn out to have very large four quark admixtures.
\end{abstract}

\maketitle

\section{Introduction}
     Recently the Belle
collaboration \cite{Choi}  provided strong 
evidence for the Z(4430) resonance in the $\Psi'$-
$\pi$ channel. It has the quantum numbers of
 $c{\bar c}n{\bar n}$ ,where $n$ stands for 
either a $u$ or $d$ quark, and would thus
seem to be a ``smoking gun" candidate for a
meson containing two quarks and two antiquarks.
Are there others? Many people believe that
the mass ordering of the candidates for 
the lightest scalar nonet suggests such a picture: 
\begin{eqnarray}
I=0: m[f_0(600)]&\approx& 500\,\,{\rm MeV} \hskip .7cm
n{\bar n}n{\bar n}
\nonumber \\
I=1/2:\hskip .7cm m[\kappa]&\approx& 800 \,\,{\rm MeV}
\hskip .7cm n{\bar n}n{\bar s}
\nonumber \\
I=0: m[f_0(980)]&\approx& 980 \,\,{\rm MeV}
\hskip .7cm n{\bar n}s{\bar s}
\nonumber \\
I=1: m[a_0(980)]&\approx& 980 \,\,{\rm MeV}
\hskip .7cm n{\bar n}s{\bar s}
\label{scalarnonet}
\end{eqnarray}
Here the postulated four quark content is displayed
for each state.
This level ordering, obtained by simply
counting the number of strange (s) type quarks,
 is seen to be 
flipped \cite{Jaffe} compared to that of the
 standard vector
meson nonet:
\begin{eqnarray}
I=1: m[\rho(776)]&\approx& 776\,\,{\rm MeV}\hskip .7cm
n{\bar n}
\nonumber \\
I=0: m[\omega(783)]&\approx& 783 \,\,{\rm MeV}\hskip .7cm
n{\bar n}
\nonumber \\
I=1/2: m[K^*(892)]&\approx& 892 \,\,{\rm MeV}\hskip .7cm
n{\bar s}
\nonumber \\
I=0: m[\phi(1020)]&\approx& 1020 \,\,{\rm MeV}\hskip .7cm
s{\bar s}
\label{vectornonet}
\end{eqnarray}
Note that the level inversion of four quark states would
hold either for ``molecular" or diquark-
 antidiquark pictures.

    There is another side to this story. Why are the
experimental candidates for a ``normal" p-wave $q{\bar q}$
scalar nonet e.g.,
\begin{equation}
     a_0(1450), K_0(1430) etc.,
\nonumber
\end{equation}
somewhat heavier than expected? A possible answer, based
on the repulsion of``two quark" and ``four quark"
states which mix with each other, was proposed some time ago
\cite{Black}, see also \cite{mixing}. This level repulsion 
also would
explain why the lower scalars seem to be unusually light.

\section{Toy model to check mixing picture}
    Note that QCD with massless light quarks obeys
SU(3)$_L$ x SU(3)$_R$ symmetry spontaneously broken
to SU(3)$_V$. We wish to realize this in a 
Lagrangian model \cite{toymodel} with linearly transforming 
chiral
nonet fields (There are necessarily two scalar nonets
as well as two pseudoscalar nonets present in this
approach). We input some physical particle masses
and predict the two vs. four quark contents of each
state. The non-inputed masses are also predicted. To
check the stability of the approach we have carried out
the calculations for the zero mass \cite{0mass} quark case, 
the
non-zero equal quark mass case \cite{equalmass} and finally 
the non equal quark
mass case. The method is conceptually straightforward
but complicated in detail.

 We employ the 3$\times$3 matrix
chiral nonet fields:
\begin{equation}
M = S +i\phi, \hskip 2cm
M^\prime = S^\prime +i\phi^\prime.
\label{sandphi}
\end{equation}
The matrices $M$ and $M'$ transform in the same way under
chiral SU(3)x SU(3) transformations but
may be distinguished by their different U(1)$_A$
transformation properties. $M$ desribes the ``bare"
 quark antiquark scalar and pseudoscalar nonet fields while
$M'$ describes ``bare" scalar and pseudoscalar fields
containing two quarks and two antiquarks. At the
symmetry level with which we are working, it is is unnecessary 
to
further specify the four quark field configuration.

The Lagrangian density is: 
\begin{eqnarray}
{\cal L} = - \frac{1}{2} {\rm Tr}
\left( \partial_\mu M \partial_\mu M^\dagger
\right) - \frac{1}{2} {\rm Tr}
\left( \partial_\mu M^\prime \partial_\mu M^{\prime \dagger} 
\right)
\nonumber \\
- V_0 \left( M, M^\prime \right) - V_{SB},
\label{mixingLsMLag}
\end{eqnarray}
where $V_0(M,M^\prime) $ stands for a function made
from SU(3)$_{\rm L} \times$ SU(3)$_{\rm R}$
(but not necessarily U(1)$_{\rm A}$) invariants
formed out of
$M$ and $M^\prime$.
  The leading choice of terms
corresponding
to eight or fewer quark plus antiquark lines
 at each effective vertex
reads:
\begin{eqnarray}
&&V_0 =-c_2 \, {\rm Tr} (MM^{\dagger}) +
c_4^a \, {\rm Tr} (MM^{\dagger}MM^{\dagger})
\nonumber \\
&&+ d_2 \,
{\rm Tr} (M^{\prime}M^{\prime\dagger})
+ e_3^a(\epsilon_{abc}\epsilon^{def}M^a_dM^b_eM'^c_f + 
h.c.)
\nonumber \\
&& +  c_3\left[ \gamma_1 {\rm ln} (\frac{{\rm det} 
M}{{\rm det}
M^{\dagger}})
+(1-\gamma_1)\frac{{\rm Tr}(MM'^\dagger)}{{\rm
Tr}(M'M^\dagger)}\right]^2.
\label{SpecLag}
\end{eqnarray}
All the terms except the last two
(more discussion of them is given in \cite{inst}) 
possess the  U(1)$_{\rm A}$
invariance.
The symmetry breaking term which models the QCD
 mass term
takes the form:
\begin{equation}
V_{SB} = - 2\, {\rm Tr} (A\, S)
\label{vsb}
\end{equation}
where $A=diag(A_1,A_2,A_3)$ is proportional to
the three light quark mass
matrix, $diag(m_u, m_d, m_s)$.
The model allows for two quark condensates,
$\alpha_a=\langle S_a^a \rangle$ as well as
four quark condensates
$\beta_a=\langle {S'}_a^a \rangle$.
These are assumed to obey isotopic spin
symmetry:
\begin{equation}
\alpha_1 = \alpha_2  \ne \alpha_3, \hskip 2cm
\beta_1 = \beta_2  \ne \beta_3
\label{ispinvac}
\end{equation}

 We also need the ``minimum" conditions,
\begin{equation}
\left< \frac{\partial V_0}{\partial S}\right> + \left< 
\frac{\partial
V_{SB}}{\partial
S}\right>=0,
\quad \quad \left< \frac{\partial V_0}{\partial S'}\right>
=0.
\label{mincond}
\end{equation}
There are twelve parameters describing the Lagrangian and the
vacuum. These include the six coupling constants
 given in Eq.(\ref{SpecLag}), the two quark mass parameters,
($A_1=A_2,A_3$) and the four vacuum parameters ($\alpha_1
=\alpha_2,\alpha_3,\beta_1=\beta_2,\beta_3$).The four minimum
equations reduce the number of needed input parameters to
eight.

Five of these eight are supplied by the following
masses together with the pion decay constant:
\begin{eqnarray}
 m[a_0(980)] &=& 984.7 \pm 1.2\, {\rm MeV}
\nonumber
\\ m[a_0(1450)] &=& 1474 \pm 19\, {\rm MeV}
\nonumber \\
 m[\pi(1300)] &=& 1300 \pm 100\, {\rm MeV}
\nonumber \\
 m_\pi &=& 137 \, {\rm MeV}
\nonumber \\
F_\pi &=& 131 \, {\rm MeV}
\label{inputs1}
\end{eqnarray}
Because $m[\pi(1300)]$ has such a large uncertainty,
we will, as previously, examine predictions
depending on the choice of this mass
within its experimental range.
The sixth input will be taken as the light
``quark mass ratio" $A_3/A_1$, which will
be varied over an appropriate range.
 The remaining two inputs will be taken from the
 masses of the four (mixing) isoscalar, pseudoscalar
mesons. This mixing is characterized by a 4x4 matrix
$M_\eta^2$. A practically convenient choice is to consider
Tr$M_\eta^2$ and det$M_\eta^2$ as the inputs.
Note that the presence of the last two terms in
Eq.(\ref{SpecLag})- which exactly mock up the QCD 
 U(1)$_A$ anomaly - decouples the initial treatment
of the other particles from that of the complicated 
pseudoscalar singlet sector.

Given these inputs there are a very large number of
predictions. At the level of the quadratic terms in the 
Lagrangian,
we predict all the remaining masses and decay constants as 
well
as the angles describing the mixing between each of
($\pi,\pi'$),
($K,K'$), ($a_0,a_0'$), ($\kappa,\kappa'$) multiplets
and each of the 4$\times$4
isosinglet mixing matrices
 (each formally described by six angles).

\section{Brief summary of results}
   A detailed report on the results when the SU(3)
flavor symmetry breaking, which yields $m_K\ne{m_\pi}$,
 is taken into account will be given elsewhere 
(in preparation). Here we just mention some main
 features.

    It is comforting to first note that the 
appropriate predicted results do not change much 
as one proceeds from zero quark masses (and hence, 
by spontaneously broken chiral symmetry, zero 
masses for the lighter pseudoscalar octet) to 
non-zero but degenerate light quark masses (and 
hence, the lighter pseudoscalar masses all taking 
the value, $m_\pi$) and finally to the realistic 
case where $m_\pi$, $m_K$ and $m_\eta$ all differ 
from each other. The zero quark mass case is an 
especially important ``touchstone" since it was noted 
in the second of ref.\cite{toymodel}
 that there are actually 
twenty one different allowed terms which might 
replace the symmetry breaker, Eq.(\ref{vsb}).
 Hence, 
without such a check, one might worry that the 
somewhat surprising results obtained could be an 
artifact of a particular choice of symmetry
 breaking terms.

   In the zero light mass case the only change
in the inputs of Eq.(\ref{inputs1}) is
 to set $m_\pi$ = 0. A suitable choice for the 
``adjustable" parameter, m[$\pi$(1300)] turned 
out \cite{0mass} to be 1215 MeV.
 Then the masses of the two 
scalar
SU(3) singlets are predicted as,
\begin{equation}
m_\sigma \approx 450 MeV \hskip .7cm 
m_{\sigma'} \approx 1500 MeV.
\label{scalarsinglets}
\end{equation}
Clearly the lighter SU(3) singlet is the 
abnormally light $f_0(600)$ candidate. These two
states turn out to be roughly the linear combinations,
\begin{equation}
\frac{1}{\sqrt{6}}[(S_1^1+S_2^2+S_3^3)\pm
({S'}_1^1+{S'}_2^2+{S'}_3^3)],
\label{5050}
\end{equation}
so the ``sigma" appears to be 50 per cent two quark 
and 50 percent four quark in nature.
As for the other SU(3) multiplets, the model 
predicts the 2 quark [2] and four quark [4]
contents as roughly :
\begin{eqnarray}
&&lighter\, 0^+\,octet:\, 0.24[2]\,\, 0.76[4]
\nonumber \\
&&lighter\, 0^-\,octet:\, 0.83[2]\,\, 0.17[4]
\label{octetcontents}
\end{eqnarray}
The difference between the mainly two quark lighter 
$0^-$ octet and the mainly four quark lighter $0^+$
octet is evident. Since, for example, the lighter and
heavier $0^+$ octets mix with each other, the heavier
$0^+$ octet would have a 24 percent four quark content
and a 76 percent two quark content.

    These results are not essentially changed in the 
case \cite{equalmass} when the light $0^-$ octet
 has the mass,
 $m_\pi$ instead of mass zero.                

    To trace what happens in the scalar isosinglet
 sector when 
the SU(3) symmetry 
breaking is turned on, we note that all four such 
states (i.e. including the appropriate SU(3) octet 
members) will mix with each other. We use the
 convenient basis fields:
\begin{equation}
 \frac{S_1^1+S_2^2}{\sqrt{2}},\,\, S_3^3,\,\,
 \frac{{S'}_1^1+{S'}_2^2}{\sqrt{2}},\,\, {S'}_3^3, 
\label{basis}
\end{equation}
and label the four scalar isosinglets, which are linear
combinations of the above, in
 order of increasing mass as $f_1,f_2,f_3,f_4$.
The lightest, $f_1$ is identified with 
the ``sigma" and is predicted to have a mass about 
730 MeV which is qualitatively similar to that of the 
previous lighter scalar SU(3) singlet state. It is
predicted to have 
percentages of the basis states in Eq.(\ref{basis})
\begin{equation}
0.38,\,\,0.06,\,\,0.32,\,\,0.24.
\label{f1details}
\end{equation}
Thus we may give the 2 quark and 4 quark percentages
of the ``sigma" as
\begin{equation}
f_1:\,0.44[2],\,\,0.56[4],
\label{f1}
\end{equation}
which is similar to the previous 50-50 split.
The predicted two quark vs. four quark percentages for
some other lighter particles in this model are,
\begin{eqnarray}
&&\pi: \,\, 0.85[2], \,\, 0.15[4]
\nonumber \\
&&K: \,\, 0.86[2], \,\, 0.14[4]
\nonumber \\
&&\kappa: \,\, 0.09[2], \,\, 0.91[4].
\label{2vs4}
\end{eqnarray}
These are again similar to those in
 Eq.(\ref{octetcontents}). It seems as though,
large four quark content for the lighter
scalar states is a stable result of the present 
model.

\section{Physical interpretation of scalar masses}
    The masses obtained above appear as tree level
quantities in the effective Lagrangian under discussion.
Especially in the case of the scalars the physical
states are rather broad and will appear as poles
 in the scattering of two pseudoscalar mesons. 
A simple way to estimate the scattering
amplitude is to first
compute the tree level scalar partial wave scattering  
amplitude and then unitarize it
 by using the K-matrix method.
This is equivalent to an earlier approach \cite{as}
 and amounts to replacing the tree level
 amplitude, $T_{tree}$ by,
\begin{equation}
T=\frac{T_{tree}}{1-iT_{tree}}.
\label{kmatrix} 
\end{equation}
For example, the sigma pole at 730 MeV
discussed in the last section appears
 in the unitarized
pion scattering amplitude at  
$z\equiv M^2 
-iM\Gamma$ with
\begin{equation}
M=473 MeV, \,\, \Gamma= 473 MeV.
\end{equation}
This is of the same order as usual
 estimates for the sigma. Such scattering
calculations should be performed to find the
``actual" mass and width parameters for
all the scalars in the present model.

\section*{Acknowledgments} 

One of us (JS) would like to thank the organizers
of QCD08 for 
arranging such a 
stimulating and friendly conference.
We are happy to thank A. Abdel-Rehim, D. Black,
 M. Harada, S.
Moussa, S. Nasri and F. Sannino
 for many helpful related
discussions. This work  
was supported in part by the U. S. DOE 
under contract no. DE-FG-02-85ER 40231.

\end{document}